\documentclass[11pt,a4paper]{emulateapj}
\pdfoutput=1 
\usepackage{hyperref}
\usepackage{amsmath,amstext}
\usepackage[T1]{fontenc}
\usepackage{apjfonts} 
\usepackage[figure,figure*]{hypcap}
\usepackage{graphicx}	
\usepackage{amsmath}	
\usepackage{amssymb}	

\newcommand{\lum}{{\rm erg\,s^{-1}}}
\newcommand{\gae}{\lower 2pt \hbox{$\, \buildrel {\scriptstyle >}\over {\scriptstyle\sim}\,$}}
\newcommand{\lae}{\lower 2pt \hbox{$\, \buildrel {\scriptstyle <}\over {\scriptstyle\sim}\,$}}

\shorttitle{Kinematic study of 3FGL J1544.6-1125}
\shortauthors{C.~T. Britt et al.}

\begin{document}

\title{Orbital Dynamics of Candidate Transitional Millisecond Pulsar 3FGL J1544.6-1125: An unusually face-on system}

\author{C. T. Britt\thanks{E-mail: britt@pa.msu.edu}}
\author{J. Strader}
\author{L. Chomiuk}
\author{E. Tremou}
\author{M. Peacock} 
\affiliation{Center for Data Intensive and Time Domain Astronomy, Department of Physics and Astronomy, Michigan State University}
\author{J. Halpern}
\affiliation{Columbia Astrophysics Laboratory, Columbia University, 550 West 120th Street, New York, NY 10027-6601, USA}
\author{R. Salinas}
\affiliation{Gemini Observatory, Casilla 603, La Serena, Chile}

\begin{abstract}
We present the orbital solution for the donor star of the candidate transitional millisecond pulsar 3FGL J1544.6-1125, currently observed as an accreting low-mass X-ray binary. The orbital period is $0.2415361(36)$ days, entirely consistent with the spectral classification of the donor star as a mid to late K dwarf. The semi-amplitude of the radial velocity curve is exceptionally low at $K_2=39.3\pm1.5$\,km\,s$^{-1}$, implying a remarkably face-on inclination in the range 5--8$^{\circ}$, depending on the neutron star and donor masses. After determining the veiling of the secondary, we derive a distance to the binary of $3.8\pm0.7$ kpc, yielding a 0.3--10 keV X-ray luminosity of $6.1\pm1.9\times10^{33}\,\lum$, similar to confirmed transitional millisecond pulsars. As face-on binaries rarely occur by chance, we discuss the possibility that \emph{Fermi}-selected samples of transitional milli-second pulsars in the sub-luminous disk state are affected by beaming. By phasing emission line strength on the spectroscopic ephemeris, we find coherent variations, and argue that these variations are most consistent with emission from an asymmetric shock originating near the inner disk.
\end{abstract}


\section{Introduction}
The existence of old millisecond pulsars (MSPs) \citep{Backer82} has been explained by the process of recycling, where a slow-spinning neutron star (NS) is spun-up by accretion of material from a donor star as a low mass X-ray binary (LMXB) \citep{Bisnovatyi74,Alpar82,Radhakrishnan82}. Eventually, the binary detaches and leaves a rapidly spinning neutron star as a rotation-powered MSP. Studies of accreting millisecond X-ray pulsars have shown a direct link between accreting LMXBs and rapidly rotating NSs \citep{Wijnands98}, and further that the neutron spin period in these systems is decreasing \citep{Hartman08,Patruno10,Riggio11}, showing that accretion does in fact spin-up neutron stars at the rates necessary for recycling to occur. 
Because radio MSPs are the end state of the binary evolution of neutron star X-ray binaries (XRB), which only last $\sim1$ gigayear in the XRB state \citep{Pfahl03}, the number of MSPs should be an order of magnitude higher than the population of XRBs \citep{Kulkarni88,Pfahl03}, such that the discovered population of MSPs represents only the tip of the iceberg. 

In recent years, the missing link in this evolution was discovered in the form of transitional millisecond pulsars (tMSPs) \citep{Archibald09,Papitto13}, which are seen to transition either between an accreting X-ray binary-like state with a sub-luminous accretion disk to a radio MSP state or vice versa. One object, PSR J1023+0038, has been seen to transition in both directions \citep{Patruno14}. In the MSP state, the radio spectrum is a characteristically steep synchrotron spectrum, while in the sub-luminous disk state, the radio emission has a flat spectrum characteristic of self-absorbed synchrotron emission \citep{Archibald09,Papitto13,Patruno14,Deller15}. To date, three tMSPs have been confirmed by observing the system in each state \citep[i.e. PSR J1023+0038, IGR J18245-2452, and XSS J12270-4859][]{Archibald09,Patruno14,Papitto13,Bassa14} Of the confirmed systems, only IGR J18245-2452 resides in a globular cluster.  

The three confirmed tMSPs share some characteristics that may or may not be coincidental. They all have orbital periods near 0.2--0.4 days, implying similarly sized donor stars; in the sub-luminous disk state, they all: 
\begin{itemize}
\item{seem to follow the same radio--X-ray relationship as black hole binaries on the way to quiescence, $L_R\propto L_X^{0.7}$, at a lower normalization \citep{Deller15}.} 
\item{have strong $\gamma$-ray emission without detected pulsations \citep{Stappers14}} 
\item{show strong optical emission lines in the sub-luminous disk state, but not in the radio MSP state \citep{Archibald09,Papitto13,Bassa14}} 
\item{show X-ray pulsations \citep{Archibald15,Jaodand16}}
\item{have an X-ray lightcurve that rapidly switches between a "high" state of $\sim10^{34}\,\lum$ and a "low" state roughly an order of magnitude fainter, with occasional flares a factor of several brighter than the high state \citep{Patruno14,Linares14a,Bogdanov15b}. }
\end{itemize}
In the radio MSP state, all three are so-called "redback" pulsars with companions $\gae 0.2$\,M$_{\odot}$ that are being ablated by the high energy emission from the pulsar \citep{Roberts13}. This ablation leads to material surrounding the binary that often eclipses the radio emission, making it difficult to detect redbacks in radio MSP searches. The GeV gamma-ray emission is not eclipsed, explaining why follow-up of Fermi sources has led to a huge increase in the number of redbacks known. It has been suggested that all redback pulsars are in fact tMSPs \citep{Linares14b}, but so far no transitions have been observed except for the three systems discussed above. As it is not yet entirely clear what drives these transitions, it is also not yet clear if the redback state may persist for some time after transitions cease. The gamma-ray emission of radio tMSPs is often pulsed, though pulsations have not been detected for all sources \citep{Roberts13}. 

Other candidate tMSPs have been identified \citep[e.g. 3FGL J1544.6-1125, 3FGL J0427.9-6704][]{Bogdanov15,Strader16} which show bright gamma-ray emission in the sub-luminous disk state with an optical accretion disk, while the $\sim200$ known LMXBs are not bright GeV gamma-ray emitters \citep{3FGL}. 

One of these candidate tMSPs, 1RXS J154439.4-112820 =  3FGL J1544.6-1125, was suggested to be a tMSP in the sub-luminous LMXB state on the basis of its X-ray and $\gamma$-ray properties \citet{Bogdanov15}. The X-ray emission shows variability which, phenomenologically, has only been seen in tMSPs in the sub-luminous LMXB state. This object has only been observed in a disk state. While the switch to or from a radio MSP state is the defining characteristic of tMSPs, there may be a decade or more between switches for a given system such that transition may simply have not occurred in 3FGL J1544.6-1125 since it was identified as a gamma-ray source by the {\em Fermi} Gamma-ray Point Source Catalog \citep{3FGL}. 

The physical mechanism behind state transitions remains poorly understood, with the predominant theory being gating of a weak accretion flow by the neutron star's magnetic field until the accretion rate rises enough to overcome the barrier \citep{Papitto14,Papitto15a}. In this picture, in the radio MSP state, the radius at which the accretion disk is truncated by the magnetic field, $R_{in}$, is outside the light cylinder, $R_{lc}=cP/2\pi$, where $c$ is the speed of light and $P$ is the spin period of the pulsar. When this condition is met, the disk is disrupted and particles are blown from the system by the pulsar wind. \citet{Papitto15a} argues that the magnetic pressure falls off with radius more slowly than the pressure from the accretion disk and the entire disk is disrupted, while \citet{Takata14} argues that some of the outer disk may remain at $R=10^{9-10}$ even if $R_{in}>R_{lc}$, with the inner regions of the disk being evaporated by gamma-ray production inside the pulsar's light-cylinder \citep[see e.g. Figure 2 of][]{Takata14}. If the accretion rate increases, the pressure from the disk can increase enough for the truncation radius to move inside $R_{lc}$ and a disk can form. However, even having $R_{in}<R_{lc}$ does not guarantee accretion onto the neutron star surface; if the magnetic field at $R_{in}$ moves faster than the Keplerian velocity, then material will still be ejected from the system. This is the so-called "propeller state" \citep[see][for a review]{Lipunov87}, in which a disk can form, but material does not reach the neutron star surface. Only if $R_{in}$ pushes in below the co-rotation radius, $R_{c}=(GMP^2/4\pi^2)^{1/3}$, can material accrete unfettered onto the NS surface. These thresholds reached by a changing accretion rate would then account for the 3 broad states observed in IGR J18245-24522: a luminous LMXB state at $L_{X}\gae10^{35}\lum$ ($R_{in}<R_{c}$), a sub-luminous LMXB state at $L_{X}\approx10^{33-34}\lum$ ($R_{c} < R_{in}<R_{lc}$), and radio MSP at $L_{X}\lae10^{31}\lum$ ($R_{in} > R_{lc}$) \citep{Papitto15a}. In the model proposed by \citet{Takata14}, the inner disk in the LMXB-like state is evaporated before reaching $R_{lc}$, and X-ray and UV photons from the hottest parts of the disk undergo inverse Compton scattering in an intrabinary shock where the pulsar wind meets the solar wind to produce the gamma-ray emission.

These state transitions are independent from the fast-switching bimodal X-ray light curves observed in confirmed tMSPs in the sub-luminous LMXB-state, where they switch on a time-scale of tens of seconds between a high-state and a low-state which is an order of magnitude fainter, with intermittent bright flares \citep{Bogdanov15,Papitto13,deMartino10,deMartino13,CotiZelati14,Bogdanov14}. X-ray pulsations are seen during the high mode, but not during the low mode or flares \citep{Archibald15}. \citet{Campana16} propose a model for the mode-switching which invokes the mass flow being briefly truncated by the light cylinder in the low modes and entering a propeller state in the high modes.

The origin of the gamma-ray emission in the sub-luminous disk state remains uncertain; either it originates from self-synchrotron Compton emission at the boundary between a disk inflow and propelling magnetosphere \citep{Papitto15a}, from a shock between the pulsar wind and mass in-flow \citep{CotiZelati14,Stappers14} or indirectly through inverse Compton scattering of disk photons in the pulsar wind \citep{Takata14}. Detections of X-ray pulsations in the sub-luminous disk state \citep{Archibald15,Papitto15b} show that at least some material is reaching the neutron star surface, which should also quench the pulsar wind, disfavoring scenarios invoking a rotation-powered pulsar wind. In the MSP state, the gamma-rays may originate entirely from the pulsar itself, with a shock producing X-rays responsible for heating the inner face of the companion star \citep{Archibald13}. 

3FGL J1544.6-1125, as mentioned above, shares the peculiar X-ray and gamma-ray properties of other tMSPs in the sub-luminous LMXB state. It also has a variable optical counterpart of $V\approx19$ \citep{Bogdanov15}. This paper presents optical spectroscopy in an effort to determine the orbital period and other properties of 3FGL J1544.6-1125.

\section{SOAR Spectroscopy and Properties of the Secondary}
\label{sec:data}

We began spectroscopy for radial velocity monitoring of 3FGL J1544.6-1125 on 16 May 2015 with the Goodman spectrograph \citep{Clemens04} on the SOAR telescope until 2 August, 2016. All spectra for this goal used a 1200 l mm$^{-1}$ grating and 1.03\arcsec\ longslit, with typical wavelength coverage of 5500--6750 \AA\ and a resolution of 1.7 \AA\ full-width at half maximum. Exposure times were typically 900 sec, with a few taken at 1200 sec in poorer conditions such as high seeing or thin cloud cover. Two to three spectra were taken back-to-back, followed by an arc lamp exposure. On some nights only one series of exposures was obtained; on others several such series were taken in succession. In particular, on 30 Apr 2016 we observed the system for over 4 hr consecutively. All spectra were reduced and extracted in the standard manner as described in \citet{Strader15}.

We chose this wavelength range primarily to cover the bright H$\alpha$ emission line and the two \ion{He}{1} emission lines at 5875 and 6678 \AA, with an initial goal 
of determining the orbital period using these lines. While these lines are strongly in emission, this effort was unsuccessful, which can be understood in retrospect as an expected consequence of the face-on inclination of the binary (see \S \ref{sec:fit}).

While the spectra are dominated by emission lines listed above, in some of the spectra weak absorption lines from the secondary are evident. These are typically only visible in the higher signal-to-noise spectra. The strongest absorption lines at the \ion{Ca}{1} lines at 6122 and 6162 \AA, though other lines were also present. This suggests a mid-to-late K classification for the secondary. 

\begin{figure*}
\centering
\includegraphics[width=0.43\textwidth]{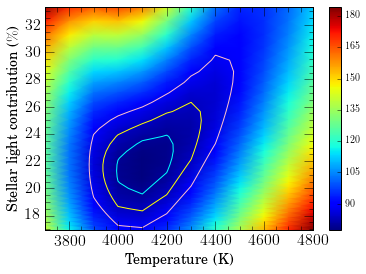}
\hspace{2em}
\includegraphics[width=0.48\textwidth]{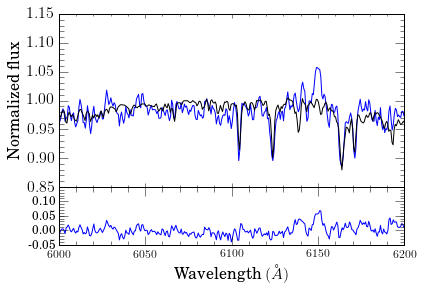}
\caption{{\em Left:} A heatmap of $\chi^2$ values for 3FGL J1544.6-1125 compared to different temperature synthetic standards masked by different disk contributions. 1, 2, and 3-$\sigma$ contours corresponding to $\Delta\chi^2_{min}+2.3,6.17,11.8$ are shown. {\em Right:} The best-fit synthetic spectrum, blurred to instrumental resolution with a veiling factor added, compared to the composite spectrum of 3FGL J1544.6-1125, with residuals shown below. The regions with faint emission lines present were not included in the fitting.
\label{fig:compspec}}
\end{figure*}

The spectral type of the donor star in 3FGL J1544.6-1125 was more precisely determined by $\chi^2$ fitting the average spectrum against synthetic standards from the Phoenix library \citep{Phoenix}. The spectral fitting was performed on the same region as the cross correlation, 6000--6200\,\AA, leaving out the region 6130--6160\,\AA\ due to contamination by faint emission lines. Much of the continuum light comes from the accretion disk and masks the strength of the absorption lines from the donor star. We accounted for this by adding a flat, featureless continuum to the synthetic spectra and fitting its normalization at the same time as the temperature. The results of this fitting are shown in Figure \ref{fig:compspec}. We find a best-fit temperature for the donor star of $T_{eff}=4100\pm100$\,K and that the donor star contributes $22\pm2\%$ of the continuum light, meaning the donor is likely a K6-7V star which has a main sequence mass of $0.7$\,M$_{\odot}$. If irradiation is heating the donor, or if the star's structure has been affected by binary interactions, the mass could be lower; indeed many redbacks have undermassive companions for their spectral type \citep{Roberts13}. We therefore adopt $M_2 \lae 0.7$\,M$_{\odot}$ as an upper limit on the donor mass. As can be seen in Figure \ref{fig:compspec}, there is some covariance between the temperature and veiling fraction at larger deviations from the minimum $\Delta\chi^2$, such that larger distances for 3FGL J1544.6-1125 are more likely that closer ones on the tail of the probability distribution, though the 1-$\sigma$ errors are very similar.

To determine the radial velocities of the secondary, we cross-correlated the 3FGL J1544.6-1125 spectra with the spectrum of a bright K4 star taken with the same setup over the wavelength range 6040--6290 \AA. We performed the cross-correlations using the FXCOR task in IRAF\footnote{PyRaf v2.2, Python v2.7, IRAF is distributed by the National Optical Astronomy Observatories, which are operated by the Association of Universities for Research in Astronomy, Inc., under cooperative agreement with the National Science Foundation.} with high-pass filtering of the spectrum and fitting a sinc function to the resulting CCF. We retained all velocity measurements for which the cross-correlation peak was well-defined and symmetric and with a relative height of at least 0.17, with the latter value determined from previous experience. This left a set of 44 radial velocity measurements, which are listed in Table \ref{tab:rv}. The respective observation times are given as Barycentric Julian Dates on the TDB system \citep{Eastman10}. The cross-correlation analysis was repeated without filtering and fitting a Gaussian model to the CCF peaks and keeping only those observations with fitted Gaussian widths less than 5 \AA, which yielded similar results.

The reddening to 3FGL J1544.6-1125 was measured by the depth of the Na D$_1$ line using the relation $\log(E(B-V))=2.47 EW(D_1)-1.76\pm0.17$ \citep{Poznanski12}. We use only the Na D$_1$ line because the D$_2$ line is contaminated by emission at \ion{He}{1} 5875. We find a value of $E(B-V)=0.15^{+0.07}_{-0.05}$, which is consistent with the maximum interstellar reddening on this line of sight, $E(B-V)=0.179\pm0.004$ \citep{Schlafly11}. In order to calculate the distance to 3FGL J1544.6-1125, we first flux calibrated spectra on photometric nights with the flux standard Feige 110 and a $3"$ slit using the tools in the ONEDSPEC package in IRAF. We measure a flux of $F_{\nu}=8.05\times10^{-28}$\,ergs\,s$^{-1}$\,cm$^{-2}$\,Hz$^{-1}$ at 6210\,\AA, and estimate an $r'$ magnitude of $19.1$ at the time of spectroscopy. Using the fit values for the veiling fraction from the spectral fit, we estimate the Vega magnitude of the donor star at this wavelength is $r'=20.7$. We used the measured $T_{eff}=4100\pm100$\,K and interpolation in the $T_{eff}-M_r$ relation from Table 5 of \citet{Kraus07} to infer the intrinsic $M_r$ of the secondary. Because the donor star may be bloated, the usual $M_r$ for this temperature may not be quite correct, but because the orbital period is so close to that expected for the spectral type from the mass-period relation found in Chapter 4 of \citet{Frank02} which is driven by the radius of the donor star, the radius of the donor is unlikely to differ by more than $\sim10\%$ from the normal main sequence size. This is because the average density of the star is set entirely by the orbital period \citep{Frank02}. In the case of XSS J12270-4859, for example, the orbital period of $\approx0.7$ hours implies a density similar to a K3V/K4V star, but the spectral type is much earlier, at F5V-G5V, depending upon phase \citep{deMartino14}. XSS J12270-4859 therefore does not fall on the Main Sequence relation and so must be undermassive. 3FGL J1544.9-1125, by contrast, falls very near to the Main Sequence relation, such that its density must closely match a normal main sequence K6V/K7V star. This means that the primary uncertainty in the absolute magnitude of the star is the temperature and veiling fraction.

Using the extinction law from \citet{Nataf13} and filter properties from \citet{Schlegel98}, the distance modulus for the donor star is then $\mu=12.9$, giving a distance of $3.8\pm0.7$\,kpc, where the primary statistical uncertainty is from the fitting the temperature and donor contribution to the continuum. This distance implies a $0.3-10$\,keV X-ray luminosity in {\em XMM-Newton} data of $6.1\pm1.9\times10^{33}\,\lum$, a gamma-ray luminosity of $(2.34\pm0.77)\times10^{34}\,\lum$, and a $0.3-79$\,keV X-ray luminosity of $(1.45\pm0.45)\times10^{34}\,\lum$ in a joint {\em XMM-NuStar} spectral fit \citep{Bogdanov16}. We note that the outer wings of the $\chi^2$ fit are not symmetric such that larger distances are more likely than smaller ones after 1-$\sigma$. The quoted error does not include possible effects from the size of the donor star being distorted from the main sequence, as the physical size of the donor is not well constrained beyond knowing the orbital period. 

\begin{deluxetable}{crr}
\tablewidth{0.4\textwidth}
\tabletypesize{\footnotesize}
\tablecolumns{3}
	\tablecaption{Radial Velocities of 3FGL J1544.6-1125 \label{tab:rv}}
	\tablehead{\colhead{BJD} & \colhead{RV} & \colhead{Err.} \\
			 \colhead{(d)}  & \colhead{(km s$^{-1}$)} & \colhead{(km s$^{-1}$)} 
			 }
	\startdata
	2457158.5964848 & 104.2 & 9.9 \\
	2457166.5953484 & 124.0 & 6.9 \\
	2457166.6060673 & 111.5 & 9.3 \\
	2457166.7633367 & 144.0 & 6.8 \\
	2457196.6330950 & 183.9 & 8.9 \\
	2457196.6671175 & 173.0 & 8.1 \\
	2457252.5499762 & 118.6 & 6.7 \\
	2457252.5637513 & 98.0 & 8.9 \\
	2457252.5744459 & 105.4 & 6.9 \\
	2457425.8170602 & 152.0 & 7.1 \\
	2457425.8517070 & 175.4 & 7.1 \\
	2457484.7321883 & 144.5 & 9.8 \\
2457484.7884300 & 191.8 & 8.0 \\
	2457484.7991899 & 181.4 & 7.5 \\
2457508.6515649 & 137.5 & 9.7 \\
2457508.6622573 & 167.2 & 9.3 \\
2457508.6750753 & 167.4 & 9.3 \\
2457508.6861669 & 172.2 & 8.1 \\
2457508.7025386 & 181.4 & 6.5 \\
2457508.7132548 & 179.6 & 10.3 \\
2457508.7261687 & 177.4 & 5.8 \\
2457508.7368855 & 173.3 & 9.9 \\
2457508.7496788 & 162.1 & 6.8 \\
2457508.7603618 & 147.0 & 6.7 \\
2457508.7876363 & 126.7 & 7.2 \\
2457508.7983329 & 122.0 & 7.8 \\
2457508.8112030 & 102.6 & 7.7 \\
2457508.8219402 & 83.1 &  6.5 \\
2457509.7578459 & 140.6 & 9.3 \\
2457509.7719966 & 113.2 & 8.0 \\
2457509.7890654 & 101.1 & 10.1 \\
2457576.5624391 & 186.4 & 8.9 \\
2457576.5862309 & 184.6 & 8.9 \\
2457598.5833796 & 170.0 & 10.3 \\
2457598.5940661 & 178.8 & 10.0 \\
2457598.6178102 & 172.5 & 9.9 \\
2457602.5682609 & 113.5 & 7.1 \\
2457602.5811867 & 95.4 & 8.9 \\
2457603.5152893 & 110.3 & 7.3 \\
2457603.5260052 & 104.5 & 6.1 \\
2457603.5389010 & 111.0 & 6.6 \\
2457603.5496177 & 123.1 & 6.7 \\
2457603.5606534 & 134.3 & 6.7 \\
2457630.5186716 & 134.2 & 8.1 \\
\enddata
\end{deluxetable}

\section{Orbital parameters of 3FGL J1544.6-1125}
\label{sec:fit}

\subsection{Radial Velocity Curve}

\begin{figure*}
\centering
\includegraphics[width=0.45\textwidth]{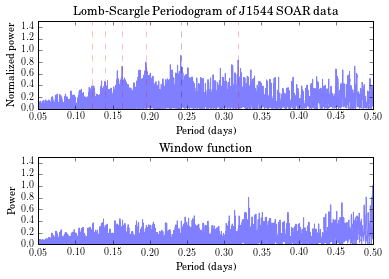}
\hspace{2em}
\includegraphics[width=0.45\textwidth]{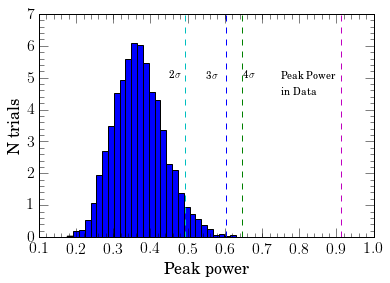}
\caption{{\em Left: }{\em top:} The normalized Lomb-Scargle periodogram of the 3FGL J1544.6-1125 RV data shows evidence of harmonics of the 1 day sampling alias (red dashed lines) $f_{alias}=f_{orbit}\pm n\times 1$\,day$^{-1}$, and a strong peak around a quarter of a day (grey dashed line) which is not present in {\em bottom: }the window function determined by the sampling pattern. {\em Right:} A normalized histogram of the peak power present in X,Y shuffling of the RV data in 121120 trials. Vertical lines show 2, 3, and 4-$\sigma$ power levels. The power in the periodogram of 3FGL J1544.6-1125 is highly significant. \label{fig:periodogram}}
\end{figure*}

An initial period search on the radial velocities in Table \ref{tab:rv} was performed using the Gatspy LombScargle package in Python \citep{VanderPlas15}. The resulting periodogram is shown in Figure \ref{fig:periodogram}, which shows a peak around $0.24$ days. To determine the significance of this peak, we used XY shuffling of the radial velocities to preserve random noise in the power spectrum and searched for periods in the shuffled data. The maximum power in each shuffled periodogram was stored for comparison to the peak power in the data. In $121120$ shuffled data sets, none had a peak power matching that found in the data itself. We ceased running simulations at this point because of diminishing returns to computational investment.

\begin{figure}
\includegraphics[width=0.45\textwidth]{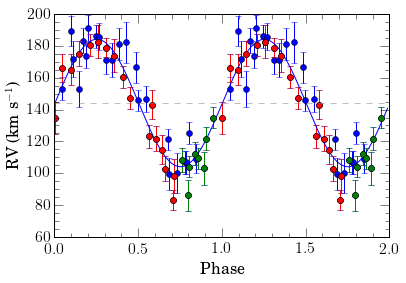}
\caption{{\em Top: } The folded radial velocity curve from the data given in Table \ref{tab:rv}, with the whole data set plotted in blue. The data from April 30th, 2016, using the ephemeris from the fit of all the data, is plotted in red, while data from August 3, 2016 is plotted in green. These nights represent the largest continuous blocks of observations with acceptable signal to noise in the spectrum.  \label{fig:rvcurve}}
\end{figure}

Further evidence that the orbital period is correct can be found in comparing the data from individual nights. In Figure \ref{fig:rvcurve}, the folded radial velocities for 3FGL J1544.6-1125 are shown alongside the night of April 30th, 2016 and August 3, 2016, with the same ephemeris applied. We use the convention where $\phi=0$ is the point where the donor star is closest to the observer. Each single night's observation is in phase with the rest of the observations and is free from the one-day aliasing that gives rise to the many peaks at integer fractions of a day in the periodogram. 

We now refine this analysis through a formal orbital fit. Given the short period and apparent active accretion, we assume a circular orbit for the binary. We fit a Keplerian model using Monte Carlo sampler \emph{TheJoker} \citep{Price-Whelan17}. The one-dimensional posterior distributions for the orbital elements are all very close to Gaussian, so we summarize the posterior values with conventional 1$\sigma$ intervals around the median values. These are: period $P = 0.2415361$(36) d, secondary semi-amplitude $K_2 = 39.3\pm1.5$ km s$^{-1}$, systemic velocity $\gamma = 143.9\pm1.2$ km s$^{-1}$,
and time of the ascending node of the neutron star $2457508.8323$(53) d.
This value is quoted as the value closest to the median epoch of the SOAR spectroscopy.

The period and $K_2$ measurement immediately give the mass function $f(M_1)$ of the primary:

\begin{equation}
f(M_1) = \frac{P K_{2}^{3}}{2 \pi G} = \frac{M_1 \, (\textrm{sin}\, i)^{3}}{(1+q)^{2}}
\end{equation}

\begin{figure}
\includegraphics[width=0.48\textwidth]{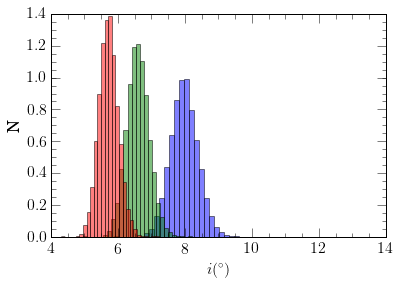}
\caption{Bootstrapping the RV measures in Table \ref{tab:rv} to fit for $P_{orb}$, $K_2$, and $\gamma$ gives the distribution of inclination angles allowed for different assumptions about the donor star and neutron star. The most conservative assumptions, with the largest binary mass ratio $q$ obtained by assuming a $1.4$ M$_{\odot}$ neutron star and a normal main sequence mass of $\sim0.7$ M$_{\odot}$ for a late K star, are shown in blue, while the effect of assuming a $2$ M$_{\odot}$ NS is shown in green. In other tMSP and redback systems, the donor stars are substantially undermassive compared to the normal main sequence mass, so that the inclination angle in 3FGL J1544.6-1125 may be even lower. Assuming an undermassive donor of $\sim0.2$ M$_{\odot}$ and a massive neutron star of $2$ M$_{\odot}$ results in the red distribution of inclination angles.  \label{fig:incs}}
\end{figure}

Here $q = M_2/M_1$ is the mass ratio of the binary and $i$ the inclination. We find $f(M_1) = 1.52^{+0.18}_{-0.17} \times 10^{-3} M_{\odot}$. Given that the spectral type of the secondary implies $M_2 \lesssim 0.7 M_{\odot}$, tight constraints on the inclination follow from an assumed primary mass. Given the compelling X-ray evidence that this system is a transitional millisecond pulsar, we assume that the primary is a neutron star in the mass range 1.4--2.0 $M_{\odot}$. This implies that $i$ must be in the range 5--8$^{\circ}$, with the lower values from massive neutron stars and low-mass secondaries, shown in Figure \ref{fig:incs}. Many redbacks and tMSPs have inflated donors, which are undermassive compared to main sequence stars of the same temperature \citep{deMartino14,Roberts13}. The less massive the donor star of 3FGL J1544.6-1125 is, and the more massive the primary neutron star is, the more face-on the system must be. In any case, 3FGL J1544.6-1125 is \emph{very} face-on.

We phase-binned the spectra using the fitted ephemeris and period in an effort to determine if the spectral type was a function of phase, which can occur when the inner face of the donor star is significantly heated. In the existing population of redback MSPs, there exist a handful of systems that do not show evidence of irradiation \citep[e.g. PSR J1628-3205][]{Li14}. We saw no significant evidence of such phase-dependent temperature changes, to a sensitivity of $\pm300$\,K, which could be a result of the face-on inclination of the system.

\subsection{Phase dependence of emission lines}

\begin{figure*}
\includegraphics[width=0.45\textwidth]{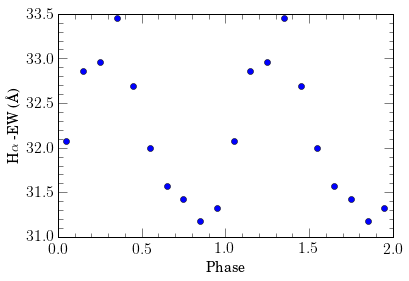}
\includegraphics[width=0.45\textwidth]{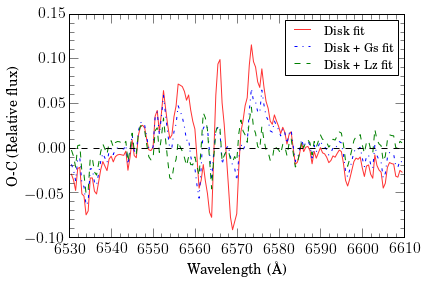}
\caption{The equivalent width of H$\alpha$ traces the optical continuum light, with a maximum at $\phi\approx 0.2-0.4$. The inner face of the donor star is most visible at $\phi=0.5$. 
{\em Right:}  The H$\alpha$ profile is best fit by the sum of a Lorentzian and a disk profile, with the accretion disk at low inclination contributing the majority of the light and the Lorentzian component, interpreted as collisional broadening in a shock, contributing $\approx10-20\%$ of the line emission. 
\label{fig:haew}}
\end{figure*}

Though the phase-binned spectra do not show any change in temperature for the donor star, the strength of the emission lines does vary with orbital phase. 
The equivalent width (EW) shows a variation at the $10\%$ level, shown in Figure \ref{fig:haew}, appearing roughly sinusoidal and peaking at $\phi\approx 0.2-0.4$. An irradiated inner face of the donor star is most visible at $\phi=0.5$, while a hotspot on the accretion disk is most visible at $\phi\approx0.75$, suggesting that the increased H$\alpha$ EW does not originate with either.

Other tMSPs at higher inclination angles have shown the characteristic double-peaked emission profile of an accretion disk. At extremely low inclination, we expect the two peaks from the receding and approaching sides of the accretion disk to merge to a single peak. We model the line profile obtained by summing all spectra to the center-of-mass frame as an optically thick, geometrically thin accretion disk following \citet{Horne86}, setting the disk emissivity profile $j(R)\propto R^{-\beta}$ to values ranging from $1.5\le\beta\le2.0$, and allow the inner disk radius, outer disk radius, and inclination angle $i$ to be free parameters. We also blur this profile by the instrumental resolution. 

A pure accretion disk fit does not provide an excellent fit to the H$\alpha$ region, with $\chi^2_{\nu}=7.27, \nu=125$, though we recover a single-peaked, thermally dominated profile as expected for a very face-on inclination. Adding a second component to account for a secondary emission site improves the fit. Fitting a joint disk plus Gaussian model gives $\chi^2_{\nu}=2.51, \nu=123$, while a better fit is obtained by using a disk plus a Lorentzian component $\chi^2_{\nu}=1.23, \nu=123$, as shown in Figure \ref{fig:haew}. A Lorentzian profile does not result from either thermal or turbulent broadening, which both produce a Gaussian profile. Instead, Lorentzian profiles of emission lines can be a result of collisional broadening, in which the width of the line is directly related to the frequency of collisions $\Gamma/\pi=\nu_{col}$, while other mechanisms such as natural broadening and the pressure broadening typical of main sequence K stars do not reach the large widths seen in these data. Based on the joint fits, the accretion disk is responsible for $80-90\%$ of the H$\alpha$ emission, while the remainder may be produced by particle collisions in a shock.
The presence of a shock contributing to the optical light would provide a source for the synchrotron emission generating the optical polarization seen in PSR J1023+0038 in the sub-luminous disk state as well \citep{Baglio16}. The fitted disk parameters are consistent with the results of the radial velocity study, with the best fit resulting in an inclination of $i=5.6^{\circ}$, inner disk radius $r_{1}\approx10^{3}$\,km, and outer disk radius $r_{2}\approx10^{6}$\,km, though inclination and outer disk radius are highly degenerate, as are, to a lesser extent, inclination and inner disk radius. The fitted value of $r_{1}$ is consistent with the model of an inner disk in between the light cylinder and co-rotation radius (i.e. in the propeller state) for a neutron star with a spin frequency of tens of Hz \citep{Papitto15a,Campana16}.

\section{Discussion}

3FGL J1544.6-1125 is remarkably face-on. Making no assumptions about the neutron star mass or donor star, it has an inclination securely below 10 degrees, with the most likely values between 5 and 8 degrees. Indeed, since these objects have likely experienced a great deal of accretion from their donor, being at the end of their lives as LMXBs according to the pulsar recycling model, the actual mass of 3FGL J1544.6-1125 may be higher than typical neutron stars, so that inclination angles towards the lower end of that range would be favored. In addition, many MSPs have donor masses well below what would be expected from a main sequence star of the same temperature, which would also imply a more face-on inclination if the same is true of 3FGL J1544.6-1125. 

Binary inclination, however, is uniformly distributed in $\cos i$ rather than in $i$ itself, and $\cos i$ at small angles is very flat and very close to unity. Given the small population, having a single gamma-ray selected tMSP with inclination below 8 degrees is somewhat unlikely to be a result of random selection from a uniform probability distribution. 

To test how unusual the known binary inclinations of gamma-ray selected tMSPSs are, we simulated data sets of 3 points drawn from a uniform probability distribution and compared them to the distribution of inclinations of the {\em gamma-ray selected} field tMSP population (shown in Table \ref{tab:msppop}). We included with the confirmed tMSP XSS J12270-4859 and 3FGL J1544.6-1125 the high-inclination candidate tMSP 1SXPS J042749.2-670434 = 3FGL J0427.9-6704 \citep{Strader16}, which is seen to eclipse in optical, X-rays, and gamma-rays, in an effort to be as inclusive and conservative as the extremely small number of known systems allows. The chances of finding at least one out of the three systems at inclinations at least as face-on as 3FGL J1544.6-1125 in a population uniformly distributed in $\cos i$ is $2.3\%$, which is suggestive but inconclusive. 

This naturally raises questions about the selection effects on the current population of tMSPs. For example, three of the five candidate or confirmed tMSPs are gamma-ray selected, running the gamut from high inclination (3FGL J0427.9-6704) to low (3FGL J1544.6-1125). If the gamma-ray emission is even slightly beamed, it would result in an overabundance of low-inclination sources in the known population. The gamma-ray luminosities, distances, and inclination angles for each system is shown in Table \ref{tab:msppop}. With a small sample size, further constrained by the ambiguous identification of gamma-ray counterparts in the crowded environments of globular clusters, it is uncertain whether any correlation between gamma-ray luminosity in the sub-luminous disk state and inclination exists. The current population of candidate and confirmed tMSPs in the Galactic field is so small as to be non-constraining. 

\begin{deluxetable*}{ccccl}
\tablewidth{0.7\textwidth}
\tabletypesize{\footnotesize}
\tablecolumns{5}
	\tablecaption{Inclinations, distances, and $L_{\gamma}$ for Redback MSPs \label{tab:msppop}}
	\tablehead{\colhead{MSP name} & \colhead{$i$} & \colhead{$d$} &  \colhead{$L_{\gamma}$}  &  \colhead{References} \\
			   & \colhead{(degrees)} & \colhead{(kpc)} & \colhead{($10^{33}$ ergs s$^{-1}$)} & 
			 }
			 \startdata
XSS J12270-4853 & $45-65$ & $1.4\pm0.2$ & $9.7\pm4.4$ & 1\\
3FGL J1544.6-1125 & $6-8$ & $3.8\pm0.7$ & $23.4\pm7.7$ & 2\\
3FGL J0427.9-6704 & $75-85$ & $2.4\pm0.3$ & $6.49\pm2.17$ & 3\\
\hline
\vspace{-6pt}
\\
PSR J1023+0038 & $30-60$ & $1.37\pm0.04$ & $1.2\pm0.3$ & 4,5 \\
PSR J1723-2837 & $30-41$ & $0.75\pm0.1$ & $2.0\pm1.8$ & 6,7 \\
PSR J0212+5320 & $>70$ & $0.8\pm0.2$ & $1.31\pm1.04$ & 8,9 \\
PSR J1628-3205 & $>55$ & $1.2\pm0.5$ & $2.09\pm1.76$ & 10 \\
PSR J1816+4510 & $>60$ & $4.5\pm1.7$ & $29.4\pm12.2$ & 11 \\
PSR J2215+5135 & $52\pm3$ & $3.0\pm1.0$ & $11.7\pm5.79$ & 12,13 \\
PSR J2129-0429 & $80\pm7$ & $1.8\pm0.2$ & $4.07\pm1.49$ & 14\\
PSR J2039.6-5618 & $49\pm1$ & $0.55\pm0.35$ & $0.619\pm0.943$ & 15 \\
PSR J2339-0533 & $57\pm1$ & $1.1\pm0.3$ & $4.35\pm3.07$ & 16 \\
PSR J1810+1744 & $48\pm7$ & $2\pm0.5$ & $12.2\pm6.18$ & 13,17 \\
\enddata
\tablenotetext{a}{The first three rows are gamma-ray selected confirmed or candidate tMSPs in the sub-luminous disk state, while rows below the horizontal line indicate redback MSPs. Only MSPs outside of globular clusters with constrained inclination angles are included. Objects were drawn from A. Patruno's list of millisecond pulsars\footnote{https://apatruno.wordpress.com/about/millisecond-pulsar-catalogue/}.}
\tablenotetext{b}{Luminosities and errors are derived from the source energy flux in the Fermi 3FGL catalog and the best available distance measure. Distance measures used, in order of preference, are trigonometric parallax, spectrophotometric distance, and distances from the dispersion measure in radio observations.}
\tablenotetext{c}{{\em References:} (1) \citet{deMartino13} (2) This work (3) \citet{Strader16} (4) \citet{Bogdanov11} (5) \citet{Deller12} (6) \citet{Bogdanov14} (7) \citet{Crawford13} (8) \citet{Li16} (9) \citet{Linares17} (10) \citet{Li14} (11) \citet{Kaplan12} (12) \citet{Hessels11} (13) \citet{Schroeder14} (14) \citet{Bellm16} (15) \citet{Salvetti15} (16) \citet{Romani11} (17) \citet{Breton13}}
\end{deluxetable*}

In addition to the small numbers, the current efforts to model the inclinations of black widow and redback MSPs could suffer the same systematic effects on inclination from an extra, variable source of light that black hole binaries do \citep[see e.g.][]{Cantrell10,Farr11,Kreidberg12}. If the amount of light from an accretion disk or shock is variable on timescales near to or less than the spacing between observations, great care must be taken to properly remove this light or the inclination angle fit by lightcurve modeling codes will be lower than the real value, driving mass estimates for the primary neutron star to higher values. In effect, the extra light not properly subtracted leads to the appearance of suppressed ellipsoidal variations, otherwise indistinguishable from lower values of $\sin i$. Since the maximum neutron star mass is of great scientific interest to the broader community because of its implications for the equation of state of neutron stars, we raise the point here that modeling inclination angles by fitting ellipsoidal modulations without coincident spectroscopic efforts to isolate the light of the donor star from other light sources could potentially lead to overestimation of neutron star mass. This is in addition to the well-recognized problems of secondary irradiation in such binaries. The evidence presented here and in \citet{Baglio16} that a shock, perhaps between infalling material and the mass outflow in the high-mode sub-luminous disk state as suggested in \citep{Campana16}, or other source of synchrotron radiation may contribute to the optical light adds an extra level of complication to the problem which is not present in black hole binaries, which already require great care to be taken when modeling the inclination angle through ellipsoidal variation. Critically, a shock does not have to dominate the optical continuum to bias mass measurements made by light curve modeling at low inclinations since $M_{NS}\propto \sin^{-3} i$.

\begin{figure}
\includegraphics[width=0.45\textwidth]{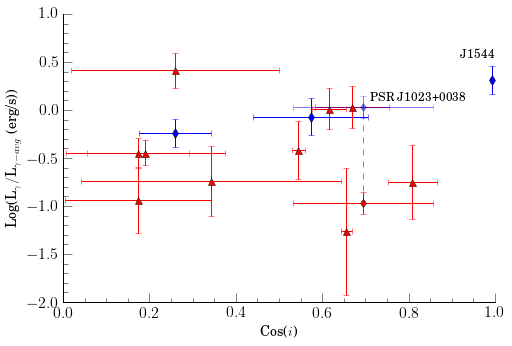}
\caption{Even slight geometric beaming could boost the gamma-ray luminosity for more face-on  systems, potentially explaining a suggested over-abundance of face-on gamma-ray selected tMSPs. tMSPs are diamonds, while tMSPs are blue.
While geometric beaming is a natural explanation for a prevalence of face-on systems, more tMSPs are needed to determine if any boosting effect is present in the sub-luminous disk state. Red points represent both redbacks (triangles) and tMSPs in the radio MSP state. There is no evidence of geometric beaming in the radio MSP state. \label{fig:beaming}}
\end{figure}

The origin of gamma-ray emission in tMSPs, and MSPs more generally, is poorly understood. Phenomenologically, the gamma-ray emission in the radio MSP state is pulsed and fainter than in the sub-luminous disk state, which has not been shown to have pulsations \citep{Johnson15}. In particular, PSR J1023+0038 brightened by a factor of several when it entered the sub-luminous disk state \citep{Takata14,Stappers14}, while XSS J12270-4859 faded by a factor of $\sim2-3$ when it left the disk state \citep{Johnson15}. LMXBs generally are not typically gamma emitters \citep{3FGL}, while tMSPs in their LMXB-like state are. Indeed, the tMSP XSS J12270-4859 was known for years as an LMXB and chosen for follow-up on the basis of its unusual gamma-ray emission \citep{Bassa14}. 

The differing gamma-ray emission between the sub-luminous disk and the MSP states observed in the known tMSPs may indicate that different mechanisms are responsible for emission in each state. It has been suggested that the gamma-ray emission in MSPs is not isotropic \citep{Romani11}, on the basis of a population of sub-luminous MSPs which have a gamma-ray luminosity below what would be expected from the spin-down luminosity compared to other, gamma-ray bright MSPs. We note that even a small Doppler boost in the direction normal to the orbital plane would make the abundance of face-on orbital inclinations much more likely in the population of gamma-ray selected tMSPs. In order to test whether there is any evidence of beaming as a selection effect in the redback population, we examine the gamma-ray luminosities of the known systems as a function of inclination angle. As shown in Figure \ref{fig:beaming}, while there may be some weak correlation between $\cos i$ and $L_{\gamma}$ among tMSPs in the sub-luminous disk state, redback MSPs in general show no such correlation. 
Given that the system geometry and gamma-ray emission is so dramatically different between states, it is plausible that geometric beaming could occur in the sub-luminous disk state but not the MSP state.

\acknowledgments
C.T.B. thanks A. Patruno for maintaining an up-to-date list of known redback MSPs in the Milky Way. JS acknowledges support from the Packard Foundation. Based on observations obtained at the Southern Astrophysical Research (SOAR) telescope, which is a joint project of the Minist\'{e}rio da Ci\^{e}ncia, Tecnologia, e Inova\c{c}\~{a}o (MCTI) da Rep\'{u}blica Federativa do Brasil, the U.S. National Optical Astronomy Observatory (NOAO), the University of North Carolina at Chapel Hill (UNC), and Michigan State University (MSU). Support from NASA grant NNX15AU83G is gratefully acknowledged.

\bibliographystyle{apj}
\bibliography{Britt}


\end{document}